\documentclass[10pt,sigconf]{acmart}

\setcopyright{none}
\usepackage[font={small,sf}]{caption}
\usepackage{setspace}
\usepackage{enumitem}
\usepackage{amsmath}
\usepackage{algorithmic}
\usepackage{algorithm}

\newcounter{myprot}
\newcounter{myalg}
\newcounter{mythm}
\newcounter{mycor}
\newcounter{myobs}
\newcounter{mydef}
\newcounter{myconj}
\usepackage{mathtools}

\usepackage{booktabs}
\usepackage{hyperref}
\usepackage{xfrac}
\usepackage{wrapfig,multirow,relsize}
\usepackage{microtype}

\newcommand{\OmitText}[1]{ {} }

\usepackage{xspace}

\newcommand{\para }[1]{\smallskip \noindent {\bf #1}}

\usepackage[all]{hypcap}
\usepackage[capitalise,nameinlink]{cleveref}
\crefname{section}{Sect.}{Sect.}
\Crefname{section}{Section}{Sections}

\usepackage{url}
\makeatletter
\g@addto@macro{\UrlBreaks}{\UrlOrds}
\makeatother

\title{Estimation of Miner Hash Rates\\ and Consensus on Blockchains}
\author{A.~Pinar Ozisik \hspace{.6cm} George Bissias \hspace{.6cm} Brian N.~Levine\\
College of Information and Computer Sciences, UMass Amherst}
\begin{document}

\setlength{\belowdisplayskip}{3pt} 
\setlength{\belowdisplayshortskip}{3pt}
\setlength{\abovedisplayskip}{3pt} 
\setlength{\abovedisplayshortskip}{3pt}

\begin{abstract}
  \em  
We make several contributions that quantify the
  real-time hash rate and therefore the consensus of a blockchain. We
  show that by using only the hash value of blocks, we can estimate
  and measure the hash rate of all miners or individual miners, with
  quantifiable accuracy. We apply our techniques to the Ethereum and
  Bitcoin blockchains;  our solution applies to any proof-of-work-based
  blockchain that relies on a numeric {\em target} for the validation
  of blocks.  We also show that if miners regularly broadcast status
  reports of their partial proof-of-work, the hash rate estimates are
  significantly more accurate at a cost of slightly higher bandwidth. Whether
  using only the blockchain, or
  the additional information in status reports, merchants can use our
  techniques to quantify in real-time the threat of double-spend
  attacks.

\end{abstract}

\maketitle
\urlstyle{sf}
\pagestyle{plain}

\section{Introduction}\label{sec:intro}
Blockchains, such as Bitcoin~\cite{Nakamoto:2009} and
Ethereum~\cite{ethereum}, are among the most successful peer-to-peer
 (p2p) systems on the Internet. Bitcoin has been adopted more widely for
e-commerce than any previous digital currency. Ethereum is also quickly
gaining prominence as a blockchain that runs Turing-complete
distributed applications. Ethereum applications include
sub-currencies~\cite{digix}, distributed autonomous
organizations with share holders~\cite{digixdao},
prediction markets~\cite{Gnosis:2016}, and games~\cite{etheria:2016};
and it currently has around half the market capitalization of Bitcoin.

There are many advantages to blockchains, including decentralized
operation. One of the primary limitations of blockchains is that the
status of transactions, blocks, and branches is not immutable --- a
fork of blocks supported by greater proof of work (POW) can emerge at any
time. Fortunately, the probability that there will be a change in
consensus regarding which fork to build on decreases
exponentially as the blockchain
grows~\cite{Nakamoto:2009}. Specifically, consider an attacker, with a
fraction of all mining power $0<q<\frac12$, that wishes to create a
fork starting from a point $z$ blocks deep on the main chain. Let the
mining power building on the main chain be $p$, such that $q+p=1$. The
probability~\cite{Feller:1968} that the attacker will eventually
produce a fork with more POW is ${(\frac{q}{p})}^z$. Several
models of blockchain security are largely based on the relative values
of $p$, $q$, and $z$; see~\cite{Garay:2015,eyal:2014}.

For merchants who are vulnerable to attack, applying
these models to a blockchain at a particular moment in time is not
possible without assigning an accurate value to $p$ or $q$. This is 
a challenging task because miners do not publish their hash rates in
real-time in a format that is verifiable by third parties. Because
no such information is available, merchants selling goods via Bitcoin
and Ethereum have no guidance for when transactions are sufficiently
safe from attacks. For example, the core Bitcoin client displays a
transaction as confirmed once it is exactly $6$ blocks deep in the
blockchain~\cite{bitcoin:confirmation}, an overly simplistic choice
that is used regardless of any other factors or conditions. Advice
from the community is appropriately vague on when a block is confirmed~\cite{Bonneau:2015a}.

To further complicate this problem, miners' hash rates can and do
change dynamically (e.g., due to diurnal electricity rates), even though the POW
algorithms are engineered to change their work targets glacially. In
other words, merchants who seek a high probability that a
transaction's status will not change should be concerned with not only
block depth, but the instantaneous hash rate of the network. Meanwhile,
honest consumers can grow impatient waiting for a large number of blocks to be appended to the blockchain
in order to get their merchandise, especially when traditional bank-based commerce
can take seconds.

\para{Contributions.} In this paper, we make several contributions
based on novel approaches to estimating the real-time hash rate of
miners and, therefore, the real-time consensus of a blockchain. We
apply our techniques to both the actual Ethereum and Bitcoin
blockchains, and in fact, they can be applied to any
proof-of-work-based blockchain that relies on a numeric {\em target}
for the validation of blocks~\cite{sasson:2014,litecoin}. Our contributions
are summarized as follows.
\begin{itemize}[nosep,leftmargin=5ex]
\item First, we design a method of accurate hash rate estimation based
  on compact {\em status reports} issued by miners. The reports add no
  computational load to miners, and are stored neither on the
  blockchain nor at peers that receive them. They can be broadcast
  off-network, for example via RSS or Twitter. Just like block
  headers, reports are verifiable as authentic POW by third
  parties.
\item Second, we show hash rates can be estimated from {\em only
    blocks} that are published to the blockchain. This approach
  requires {\em no cooperation from the miners}, but is less accurate
  than status reports.

\item We evaluate the accuracy of the two approaches using a synthetic
  blockchain as ground truth. Additionally, we derive Chernoff bounds
  for the accuracy of our estimates from status reports. We show that both methods can be used together in
  support of incremental adoption by mining pools. Our blockchain-only
  method incurs no network costs. Status reports incur extra traffic
  depending on their rate. For example, if all active miners issue about 10 reports per block, the cost is about 0.03~KBps 
  for Bitcoin and 6.6~KBps for Ethereum; note that none of the report data needs to be archived. The accuracy of both methods is tunable.

\item We apply our  estimates to calculating the probability of a double-spend attack against blocks as they garner descendants.  We show that using status reports,  99\% of blocks have a risk of 0.1\% by depth of 13 for a worst-case estimate. We show that without status reports, merchants should wait much longer. We  characterize the historical performance of Ethereum
  and Bitcoin, and show half of  blocks require a depth of at least 40 before an  attacker's
  success is below 0.1\% for a worst case estimate.  We also consider attacks against our approach. A public demo is at \url{http://cs.umass.edu/~brian/blockchain.html} that  
  provides a quantified, realtime estimate of the security of blocks. 
\end{itemize}
We begin with a background on blockchains and conclude with a discussion of limitations and related work.

\section{Background}\label{sec:basics}
The \emph{blockchain} concept was introduced by Nakamoto~\cite{Nakamoto:2009} as a
method of probabilistic distributed consensus~\cite{Vukolic:2015}.
Originally designed to be the backbone of the Bitcoin distributed
cryptographic currency, blockchains have since been applied to a
variety of scenarios. Bitcoin itself includes a scripting language
that supports a limited set of custom smart contracts.
Ethereum\cite{ethereum} is a blockchain-based distributed system that
supports Turing-complete software and includes a global data store.
Transactions in Ethereum represent the transfer of money or
data among programs, allowing for a richer space of distributed
applications. Other blockchains have been proposed and implemented to
support anonymous transactions~\cite{sasson:2014}, hedge
funds~\cite{digixdao}, medical records~\cite{Azaria:2016}, and
alternate currencies~\cite{litecoin,digix}. Below we describe the
aspects of Bitcoin and Ethereum that are relevant to us.
Tschorsch et al.~\cite{Tschorsch:2016}, Bonneau et
al.~\cite{bonneau:2015}, and Croman et al.~\cite{Croman:2016} offer
summaries of broader blockchain research issues.

\subsection{Bitcoin}
\para{Accounts.} Bitcoin users store bitcoins in accounts called {\em
  addresses}, which are created with an empty balance simply by
generating a public/private key pair. The transfer of coin between
addresses, via {\em transactions}, is recorded on a public ledger
called a {\em blockchain}. Transactions are authenticated via a
 private key signature, and the balance of each account can be 
derived from the blockchain. 

\para{Adding to the blockchain.} To be added to the blockchain,
transactions are broadcast by users on Bitcoin's p2p network. A set of
\emph{miners} on the p2p network verify that each transaction is
signed correctly, does not conflict with a previous transaction, does not
move more coin than is contained in the address, and other functions.
Each miner independently agglomerates a set of valid transactions into
a candidate \emph{block} and attempts to solve a predefined
cryptographic puzzle as POW, which involves data
from the candidate block and a specific {\em prior block}. The
new transactions  are only valid if they do not
conflict with the set of transactions that are contained in all
blocks that are direct ancestors.

The first miner to solve the problem broadcasts his solution to the
network, and by virtue of the solution, is able to add the block to
the ever-growing blockchain as a child of the prior block. The miners
then start over, using the newly appended blockchain and the set of
remaining transactions. The miners' incentive for {\em discovering}
a new block is a reward of coins, called the {\em coinbase},
consisting of a predetermined {\em block reward} (currently worth 12.5 Bitcoins)
and fees from transactions included in the block.

In Bitcoin, the POW computation is dynamically calibrated to take
approximately ten minutes per block. When transactions appear in a
block, they are \emph{confirmed}, and each subsequent block
provides additional confirmation. To announce a new block, a miner
lists all transactions contained in the new block along with a header
that contains an easily-verifiable POW solution. When a node or
miner receives a new block, he validates each transaction in the block
and the POW.

Notably, if there is a fork on the chain, honest miners always select the prior block as the last block containing the largest amount of POW. However,
due to propagation delays in the network, 
miners can receive competing (but valid) block announcements, which
bifurcates the chain, until one of the two forks is appended to first.

\para{Block Headers.} 
Any entity can elect to be a miner for Bitcoin, and there is no
centralized party from whom to seek approval for mining. If all miners
were to simply vote on which block should be appended to the main chain, then the mining process would fall vulnerable to a Sybil
attack~\cite{Douceur:2002}. The POW puzzle addresses this
problem by performing a kind of decentralized leader election: the
miner that solves the puzzle can decide which block to append to the
chain.

\para{Proof-of-Work.} 
Bitcoin uses a simple POW algorithm based on cryptographic
hashing, proposed earlier by Douceur~\cite{Douceur:2002}.
Specifically, miners apply a 256-bit cryptographic hash algorithm~\cite{hashcash} to
an 80-byte {\em block header}, and the puzzle is solved if the
resulting value is less than a known {\em target}, $0<t<2^{256}$. The
header in Bitcoin consists of the Merkle root of the set of
transactions, a timestamp, the target (stored as $2^{224}/t$), a {\em
  nonce}, and the hash of the prior block's header. If the hash is not
less than the target, then a new nonce is selected to generate a new
hash (the Merkle root can be adjusted as well). This process repeats
until some miner finds a solution.
  
Each time a nonce is selected and the block
header is hashed, the miner is sampling a value from a discrete uniform
distribution with range $[0,2^{256}-1]$. The probability of solving the POW and
discovering a block is the cumulative probability of selecting a value
from $[0,t]$, which is $t/{2^{256}}$. Hence, in expectation, the
number of samples needed to discover a block is ${2^{256}/t}$. Bitcoin
adjusts the target so that on average it takes about 600 seconds to
find a block. Typically, the target is described for convenience as a
{\em difficulty}, defined to be $D=2^{224}/t$. Bitcoin's difficulty
is set once every two weeks.

\subsection{Ethereum} Ethereum operates very similarly to Bitcoin. The
following differences are relevant to the context of this paper.
Ethereum miners solve a POW problem that is more complicated
than Bitcoin in an attempt to disadvantage miners with custom ASICs.
However, in the end, a miner still compares a hash value to the target.
Specifically, the number of values in the block header is larger,
resulting in a 508-byte header. It's not the hash of the header that
is compared against the target, but the hash resulting from an
Ethereum-specific algorithm called ETHASH~\cite{ETHASH}, for which the
hash of the block header is the primary input. In the end, the POW
hash value is a sample from a discrete uniform distribution with range
$[0,2^{256}-1]$, and the probability of block discovery is
${t/2^{256}}$.

A major difference of Ethereum is that the target is set such that the
expected time between blocks is 15 seconds. This setting results
in quicker confirmation times, but as a result, the probability that two
miners announce blocks within the propagation time of a block
announcement is much higher. Therefore, there are many abandoned
forks in the chain. Ethereum uses a modified version of the GHOST~\cite{Sompolinsky:2015}
protocol for selecting the main fork of the blockchain: the main chain
follows the block at each level with the most POW on its subtree.
These differences do not affect the application of our algorithms; in
fact, the presence of ommers is additional data which improves our
estimates.

\subsection{Double-spend Attacks}
\label{sec:ds}

The fundamental blockchain double-spend attack~\cite{Nakamoto:2009}
operates as follows. An attacker offers a transaction to a merchant in
exchange for goods. The transaction appears on the blockchain in block
$B_1$, with block $B_0$ denoting the prior block. The merchant
releases goods purchased by the transaction to the attacker only after
blocks $B_2,\dots,B_z$ follow. Honest miners, with power $0<p<1$ add
these $z-1$ new blocks. The attacker, with mining power $q=1-p$ races to add
a distinct sequence of blocks of length at least $z+1$ that forks from
$B_0$. Nakamoto derived the probability of the attacker's success of creating a longer fork, given infinite time. Nakamoto assumes that the miner's power
is constant and that she never gives up on the attack. The attacker
succeeds by producing any chain of length $z+1$ or longer, but cannot
announce the chain before the honest miners produce $B_z$ since the
merchant will not release purchased goods until then. This probability, where  $\lambda=\frac{zq}{p}$, 
is \begin{align}
  \mathcal{D}(q,z)=\begin{cases} 
 1- \mathlarger{\sum}_{k=0}^{z} \frac{\lambda^k e^{-\lambda}}{k!}\left(1-(\frac{q}{p})^{z-k}\right)
   & \text{, if }  q<\frac12 \\ 
    1 &  \text{, if } q\geq\frac12 \\
  \end{cases}\label{eq:satoshi}
 \end{align}

When $z=6$, an attacker with $q<0.127$ has a less than $0.001$
probability of success. Based on that value, the core bitcoind client
displays transactions as confirmed once they appear in a block that is
6 deep in the chain.

\section{Problem Statement}\label{sec:problem}

We seek to quantify the hash rate of miners in real-time, thereby
quantifying the consensus of a blockchain towards its blocks and the
transactions they hold.  Our goal is to increase transparency so that
merchants can judge their vulnerability to a double-spend attack; in
short, we wish to answer the question, \textit{what is the risk of releasing
goods to a consumer for a transaction that is $z$ block deep given the
current hash rate of the network?}

Specifically, in Section~\ref{sec:statusreports}, we propose that
miners periodically issue status reports that are block headers with
partial POW. Each report is exactly a block header except
the nonce corresponds to the smallest hash value the miner found
during the last $\sigma$ seconds.  We then answer these
 questions:

\medskip
 \centerline{\parbox{.8\columnwidth}{\textit{1. Given a set of status reports produced by a miner during
    a window of time, how many hashes (samples) were taken to produce
    the reports?}}}
\medskip

\noindent We also quantify the network bandwidth costs for applying status
reports to Bitcoin and Ethereum.
 
In Section~\ref{sec:mmconsensus}, we ask how we can accomplish the
task of estimating hash rate without status reports.

\medskip
\centerline{\parbox{.9\columnwidth}{
  \textit{2. Given the set of valid blocks that were added to the
    blockchain during a window of time, how many hashes (samples) were
    taken by all miners to produce the blocks?}}}\medskip

\noindent We also ask how this approach can be combined with status reports for
incremental deployment. In Section~\ref{sec:bounds}, we
characterize the accuracy of our two estimators.

In Section~\ref{sec:implementation-analysis}, we apply our 
estimators to decide when to consider a transaction
secure from double-spend attacks. Let block $B_1$ contain a transaction
of interest, and let block $B_0$ be its parent. Let $B_i$ be the $i$th
descendant from $B_0$ via $B_1$. We then estimate the hash rate of the
network in a window prior to $B_0$, and the hash rate of the network in
a window from $B_0$ to $B_i$, and we ask:

\medskip
\centerline{\parbox{.9\columnwidth}{\textit{3. Assuming the difference in hash rate is being applied
    to a double-spend attack of the transaction based on a fork from
    $B_0$, what is the probability the attacker will succeed?}}}
\medskip

\noindent We apply our algorithm to the real Bitcoin and Ethereum blockchains to
characterize the typical delay required to ensure that the risk of a double-spend
is suitably low.

\begin{figure}[t]
\hspace{-2.5ex}\colorbox{gray!10}{\relsize{-.5}
\begin{tabular}{rp{.67\columnwidth}}

\textbf{Variable} & \textbf{Description} \\
$\hat{h}$ & estimation of network hash rate \\
$\hat{h}_m$ & estimation of an individual miner's  hash rate \\
$\sigma$ & 
duration of time spent mining during interval $I$ \\
$\theta_m$ & 
number of hashes performed \textit{per miner} per status report\\
$\theta_\sigma$ & 
number of \textit{network-wide} hashes performed during time duration $\sigma$\\
$\theta$ & 
number of hashes performed, i.e. shorthand for $\theta_m$ or $\theta_\sigma$ \\
$S$      &
the size of the hash space, i.e. $S=2^{256}-1$\\
$V$      & for an arbitrary miner, a random variable representing the first order statistic (minimum hash value) after hashing $\theta$ times, where $V_i\sim Expon(\beta)$\\
$\beta$      & exponential survival parameter of $V_i$ \\
$\textbf{V}=V_1,\dots,V_n$ &
random sample of $n$ first order statistics\\
$\overline{V}$ & the \textit{observed} sample mean of \textbf{V} \\
$\hat{\beta}$ & estimation of $\beta$, using status reports\\
\vspace{3ex}
$\hat{\theta}$ &
\textit{estimated} $\theta$ from the sample population \\
$Y$ & random variable equal to 0 if no block is produced during interval $I$; else the hash of the block. Note that $Y_i$ is a function of $V_i$.\\
$\textbf{Y} = Y_1,\ldots, Y_n$ & random sample of $n$ consecutive intervals, where a block is observed or not\\ 
$\overline{Y}$ & the \textit{observed} sample mean of \textbf{Y} \\
$E[\textbf{Y}]_\beta$ &
the expected value of \textbf{Y} parameterized by $\beta$\\
$\tilde{\beta}$ & estimation of $\beta$, using only the blockchain\\
\end{tabular}
}
\caption{List of variables presented in Section~\ref{sec:statusreports} and~\ref{sec:mmconsensus}.}

\end{figure}

\section{Status Reports}
\label{sec:statusreports}
In this section, we propose that active mining pools issue {\em status reports}
periodically, which allow third parties to estimate their hash rate and to learn which specific block they are building on top of. Each report is exactly a block header except that the POW does not satisfy the current target. Instead, the minimum hash value in the report represents the hash found since the last block broadcast on
the chain or their last report. To be clear, each status report does not directly report the minimum hash value; instead, reports are of the input values to the POW algorithm. 

Below, we present and evaluate the accuracy of a method for estimating the hash
rate of each miner using status reports. In the next section, we present a technique to make estimates from only blocks on the blockchain.

\subsection{Hash Rate Estimation}
\label{sec:statusconsensus}
As described in Section~\ref{sec:basics}, the result of Bitcoin's and Ethereum's POW algorithms
is a sample taken
randomly from a discrete uniform distribution that ranges between
$[0,2^{b}-1]$, where $b=256$. The winning miner is the one that first produces a hash smaller than the target.

If a miner $m$ periodically announces the smallest value he has
discovered, we can estimate the hash rate, $\hat{h}_m$, he is lending
toward finding a successor to a given block. 
In the next section,
we detail how to accurately estimate the hash rate of cooperative
miners using status reports.

\para{Continuous model.} In order to estimate a miner's hash rate,
first we present a continuous model to describe the distribution of
the smallest hash value he reports. Let $I$  be an interval of
$\sigma$ seconds during which a miner attempts to mine a block. During
that interval, the miner hashes the block $\theta_m$ times generating
a sequence of hash values $\textbf{Z} = Z_1, \ldots,
Z_{\theta_m}$. Although we know $\sigma$, we do not know $\theta_m$
\emph{a priori}. At the end of the interval, the miner sends status
report $V = Z_{(1)}$, which denotes the lowest hash value achieved on the block
during $I$. In other words, $V$ is the random variable representing
the first order statistic after hashing the block $\theta_m$ times. Our goal
is to determine $\theta_m$ from a sample of first order statistics.

Let $S = 2^{256} -1$ be the size of the hash space. The probability
that any $V$ is greater than some $v \in [0, ~S]$ is equal to
$1 - v/S$. Thus, the probability that $V$ is always greater than $v$
across $\theta_m$ independent trials computed during ~$I$~ is given by
$(1-v/S)^{\theta_m}$. Now define $F_{V}(v~|~S)$ to be the CDF of $V$. It
follows that\vspace{-1em}
\begin{equation}
  F_{V}(v~|~S) = 1 - \left( 1 - \frac{v}{S} \right)^{\theta_m}.
\end{equation} 
Parameter $\theta_m$ can be expressed as a function of $S$:
\begin{align}
\theta_m = S / \beta.
\end{align}
Consider how $F_{V}(v~|~S)$ changes as $S$
increases. A common limit shows that
\begin{equation}
  \lim_{S \rightarrow \infty} F_{V}(v~|~S) = 1 - e^{\frac{-v}{\beta}}.
\end{equation}
Thus, $V \sim \texttt{Expon}(\beta)$, where $\beta$ is the exponential
\emph{survival parameter}. Since the mean of the exponential
distribution is equal to the survival parameter, we can see that the
expected value of $V$ is $\beta = S/\theta_m$.
We next show how status reports spanning multiple intervals can be
used to derive a robust estimator of $\theta_m$, and ultimately
$\hat{h}_m$.

\para{Estimator for hash rate.} Let $\textbf{V} = V_1, \ldots, V_{n}$
with $V_i \sim \texttt{Expon}(\beta)$ be a random sample representing
$n$ status reports, each sent $\sigma$ seconds apart by a
miner. Because each $V_i$ is $\texttt{Expon}(\beta)$ and i.i.d., the
maximum likelihood estimator of $\beta$ is equal to the sample mean,
$\overline{V}$. Furthermore, the sample mean is an unbiased estimator
of the population mean; so $\overline{V}$ is an unbiased estimator of
$\beta$.

Since $\beta = S/\theta_m$, it follows then that a reasonable
estimator for $\theta_m$ is given by\vspace{-1em}
\begin{equation}
\label{eq:theta_hat_m}
  \hat{\theta}_m = \frac{S}{\overline{V}}.
\end{equation}
Finally, the miner's hash rate $\hat{h}_m$ can be estimated as
\begin{equation}
\label{eqn:h_hat_m}
\hat{h}_m = \frac{\hat{\theta}_m}{\sigma} =  \frac{S}{\overline{V} \sigma}.
\end{equation}

\begin{figure}[t]   \centering
  \centerline{\includegraphics[width=.8\columnwidth]{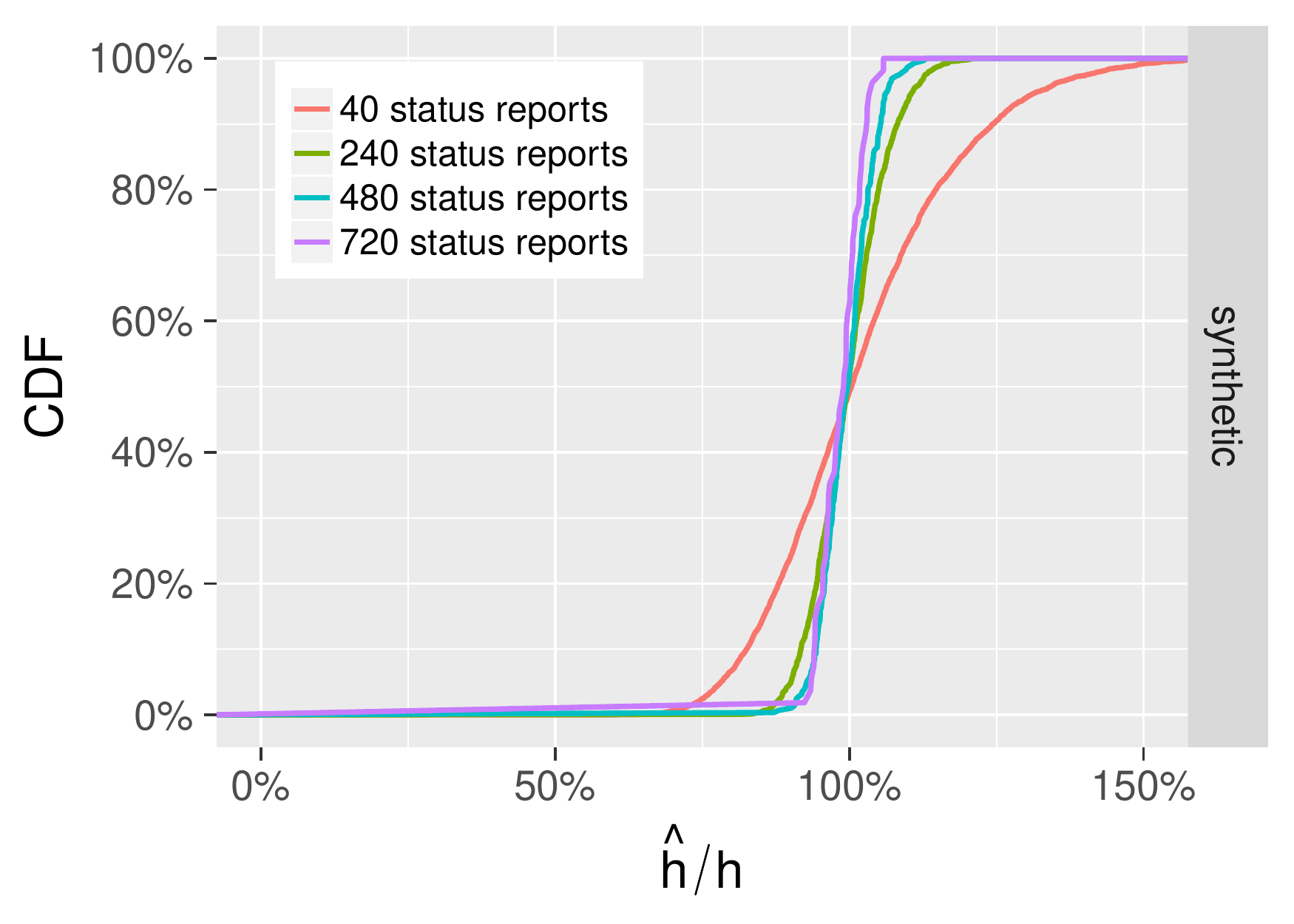}}
  \caption{A CDF of the percentage of the estimated hash rate to the
    real hash rate, as the number of status reports increase. 100\% on
    the $x$-axis denotes the line along which the estimated hash rate
    and real hash rate are equivalent.}
   \label{fig:n-hat}
\end{figure}

\para{Empirical evaluation of accuracy.} We implemented and evaluated this technique against a simulated miner. The miner sampled from a discrete uniform distribution at a fixed rate of $h$ and issued status reports regularly. Our simulated recipient estimated the hash rate, $\hat{h}$, using Equation~\ref{eqn:h_hat_m} given 40, 240, 480, or 720 reports. We measure the accuracy as $\hat{h}/h$ and show the result of tens of thousands of trials in Figure~\ref{fig:n-hat} as a CDF of the accuracy. A perfect estimate is at 100\%. 
As the number of status reports increases, a greater fraction of
our estimates are aligned with the real hash rate. More reports are required for high accuracy; however, when we apply this technique to a blockchain, we are interested in a window of blocks rather than a single block. Hence, the count of available status reports adds up quickly. 

\section{Blockchain-Only Estimation}\label{sec:mmconsensus}

In this section, we describe a blockchain-only method of estimating
miner hash rates. For this estimator, we treat the entire network as a single miner and a block as a status report 
that can only be observed at certain intervals. Although this approach has no additional network costs and does not require cooperation from miners, it is less accurate than status reports.  We then extend the technique to allow for hash rate estimation of an individual or a subset of miners. As we show, this extension allows for the incremental deployment of status reports. 

\subsection{Network Hash Rate Estimation}\label{sec:networkhash}

We would like to estimate the network hash rate, $\hat{h}$, using only
the \emph{observed} POW hash values that are represented by mined blocks. A
critical step in this process is estimating the number of hashes that
were performed network-wide using these observed values. Conceptually,
we are treating the entire network as a single miner from whom we
receive reports only when a block is mined.

Consider a window of time during which at least one block is announced. We  segment time into $\sigma$-second intervals:
$\mathcal{I} = I_1, \ldots, I_n$, and let
$\textbf{V} = V_1, \ldots, V_n$ be the minimum hash value achieved
across the entire network for each interval. Note that a valid
$V_i$ is produced during every interval $I_i$, even though it is not
broadcast to the network unless it is below the target. This notation is consistent with our
definitions from Section~\ref{sec:statusconsensus}.

Suppose that blocks were mined at the end of
intervals $\mathcal{I}_B \subseteq \mathcal{I}$ so that the observed
hash values are given by the set
$\textbf{O} = \{V_i ~|~ I_i \in \mathcal{I}_B\}$. In practice, a block
could have been mined at any point during an interval in
$\mathcal{I}_B$, but the distinction becomes less important as
$\sigma \rightarrow 0$. Finally, assume that, network-wide,
$\theta_\sigma$ hashes were performed during each interval. Our goal
is to determine an estimator of $\theta_\sigma$.

\para{Estimator for $\boldsymbol{\beta}$, the expected minimum POW hash.} In
Section~\ref{sec:statusconsensus}, we showed that
$V_i \sim \texttt{Expon}(\beta)$ as $S \rightarrow \infty$ and argued
that $\overline{V}$ is a good estimator of $\beta$. But that approach
does not work here since we are missing most of the values $V_i$.
Consider instead a new sequence of random variables
$\textbf{Y} = Y_1, \ldots, Y_n$ defined as a function of $\textbf{V}$:
\begin{equation} Y_i = \textbf{1}_{V_i \leq t}(V_i) V_i = \left \{
    \begin{array}{lr}
      V_i, & V_i \leq t \\
      0, & \text{otherwise} \end{array} \right . , 
\end{equation} 
where
$t$ is the target and $\textbf{1}_C (x)$ is the indicator
function equal to 1 when $x \in C$ and 0 otherwise. Since each $Y_i$
is a function of $V_i$ and $V_i \sim \texttt{Expon}(\beta)$, it is
straightforward to derive the expected value of any given $Y_i$:
\begin{eqnarray}
  E[Y_i]_{\beta} &=& \frac{1}{\beta} \int_0^t v e^{-v/\beta} dv \\
                 &=& \beta - \beta e^{-t/\beta}\left(\frac{t}{\beta} +
                     1\right)\label{eqn:y_ev}
\end{eqnarray}

\noindent The sample mean $\overline{Y}$ is simpler to determine. It is
the sum of the observed hash values $O_i$ divided by the number of
intervals:\vspace{-1em}
\begin{equation}
\overline{Y}=\frac{\sum_i O_i }{|\mathcal{I}|}.
\end{equation} 
Thus, we can derive the {\em method of moments} (MoM)~\cite{Casella:2002} estimator $\tilde{\beta}$ by
equating $E[Y_i]_{\beta}$ and $\overline{Y}$:
\begin{equation}
\label{eqn:mm}
\tilde{\beta} - \tilde{\beta} e^{-t/\tilde{\beta}}\left(\frac{t}{\tilde{\beta}}  + 1\right) = \overline{Y}.
\end{equation}

Unfortunately, it is difficult to solve Equation~\ref{eqn:mm} for
$\tilde{\beta}$ analytically. Moreover, the implicitly defined
$\tilde{\beta}$ is not actually a function because
$E[Y_i]_{\tilde{\beta}}$ not a one-to-one function of $\tilde{\beta}$.
And so we note that\vspace{-1ex}
\begin{equation}
\frac{\partial^2 E[Y_i]_{\tilde{\beta}}}{\partial \tilde{\beta}^2} = \frac{t^2}{b^4} e^{-t / \tilde{\beta}}(\tilde{\beta} - t)
\end{equation}
with roots at $t = 0$ and $t = \tilde{\beta}$. Because we assume that
the target $t$ is always positive, this means that
$E[Y_i]_{\tilde{\beta}}$ has a single inflection point at
$t = \tilde{\beta}$. In light of the fact that
$E[Y_i]_{\tilde{\beta}}$ is monotonic for values of $\tilde{\beta}$ on
either side of $t$, it is also possible to verify that
$E[Y_i]_{\tilde{\beta}}$ is strictly increasing for
$\tilde{\beta} < t$ and decreasing for $\tilde{\beta} > t$. Thus,
Equation~\ref{eqn:mm} implicitly defines two different functions for
$\tilde{\beta}$: $\tilde{\beta}(\overline{Y})^-$ when
$\tilde{\beta} < t$ and $\tilde{\beta}(\overline{Y})^+$ when
$\tilde{\beta} > t$.

Because both sides of the function $\tilde{\beta}(\overline{Y})$ are
monotonic, it is straightforward to solve each using binary search.
Algorithm~\ref{alg:find_beta} defines a procedure for finding
$\tilde{\beta}(\overline{Y})^+$, and $\tilde{\beta}(\overline{Y})^-$
can be found in a similar fashion.

A  drawback of the estimator is that it forces the practitioner
to guess if $\tilde{\beta}$ is greater or less than $t$ in order to
find its actual value. We find that in practice this is rarely an
issue. Recall that $\beta$ is the expected minimum hash value for a $\sigma$-second time interval, while $t$ is the target minimum hash for the much longer block creation interval. Thus, unless the mining pool under consideration is mining at many times the current difficulty, we can be quite certain that $\beta \gg t$. Accordingly, it is also likely that $\tilde{\beta}$ will also be significantly larger than $t$, which makes $\tilde{\beta}(\overline{Y})^+$ the correct branch to use. 

\begin{algorithm}
\caption{\textbf{Find }$\tilde{\beta}$}
\label{alg:find_beta}
\begin{algorithmic}[1]
\STATE Let $\tau > 0$ be some small tolerance parameter
\STATE Set $L = 0$, $H = S$, and choose $\theta_\sigma = S/2$
\STATE Set $\tilde{\beta} = S/\theta_\sigma$
\STATE Calculate $E[Y]_{\beta}$ by substituting beta tilde as beta into Equation~\ref{eqn:y_ev}
\IF{$E[Y]_{\tilde{\beta}} - \tau> \overline{Y}$} 
	\STATE Set $L = \theta_\sigma$, $H = H$, and $\theta_\sigma = L+(H-L)/2$
	\STATE Goto 3
\ELSIF{$E[Y]_{\tilde{\beta}} + \tau < \overline{Y}$}  
	\STATE Set $L = L$, $H = \theta_\sigma$, and $\theta_\sigma = L+(H-L)/2$ 
	\STATE Goto 3
\ELSE 
	\RETURN $\tilde{\beta}$
\ENDIF
\end{algorithmic}
\end{algorithm}

\para{Estimating hash rate from $\boldsymbol{\tilde{\beta}}$.} Because
$\beta = S / \theta_\sigma$, a good estimator for $\theta_\sigma$ is
$\hat{\theta}_\sigma = S / \tilde{\beta}$. This estimates the number
of hashes per $\sigma$-second interval, which we can use to
estimate the hash rate of the entire block creation interval as 
\begin{equation} \label{eqn:h_hat} 
\hat{h} = \frac{S }{\tilde{\beta}  \sigma}. 
\end{equation}

\begin{figure}[t] 
   \centering
   \includegraphics[width=.8\columnwidth]{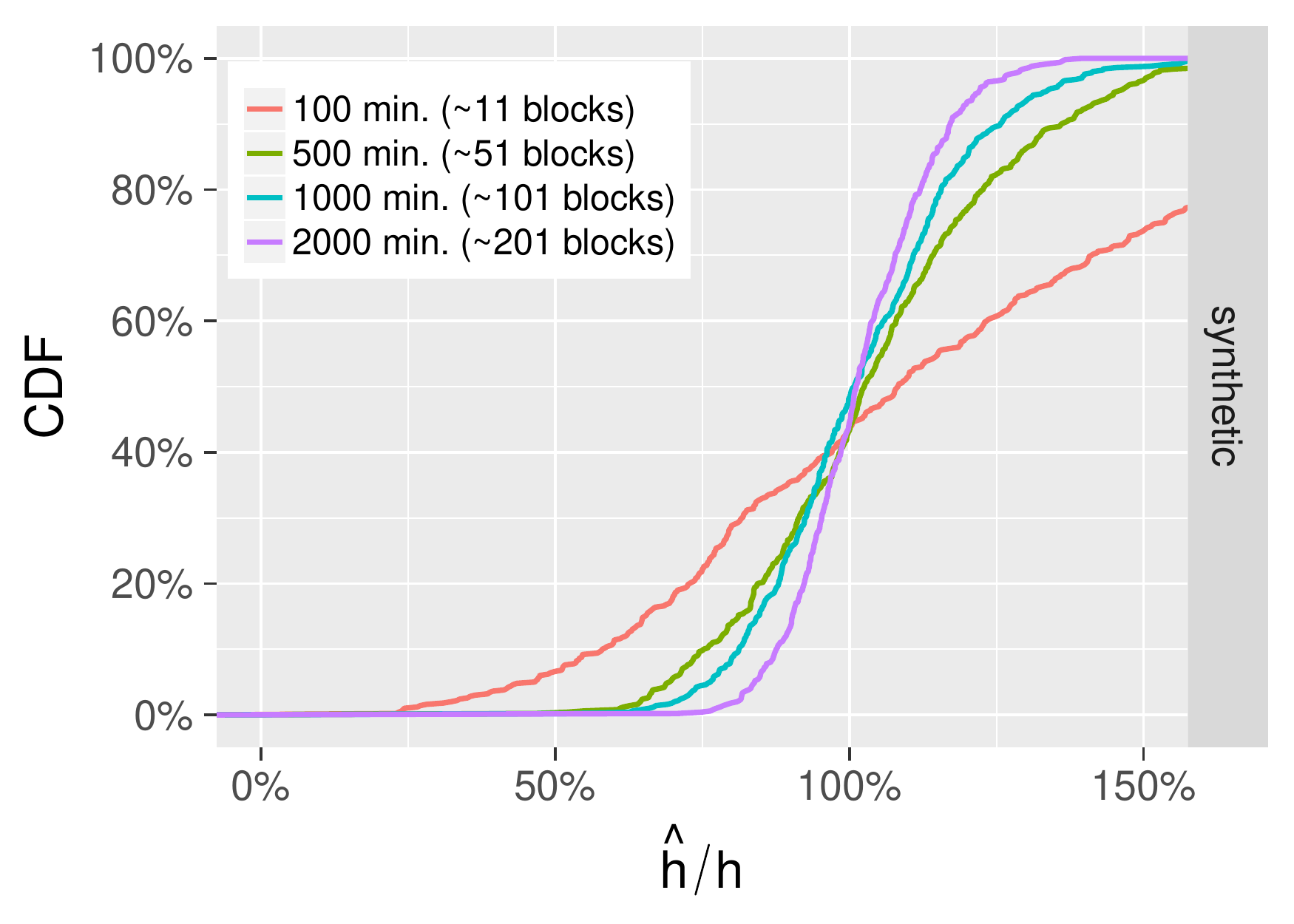} 
   \caption{A CDF of network hash rate using method of moments on a synthetic blockchain.}
   \label{fig:hashes-synthetic}
\end{figure}

\para{Empirical evaluation of accuracy.} We used the technique
described above to estimate the number of network-wide hashes on a
synthetic blockchain, using a  window
of  100, 500, 1000, or 5000 seconds,  as shown in
Figure~\ref{fig:hashes-synthetic}. As the window length increases, we
incorporate  data from additional blocks, allowing for a greater portion of
estimates to be equal to the the real hash rate.

\begin{figure}[t] 
   \centering
   \includegraphics[width=.8\columnwidth]{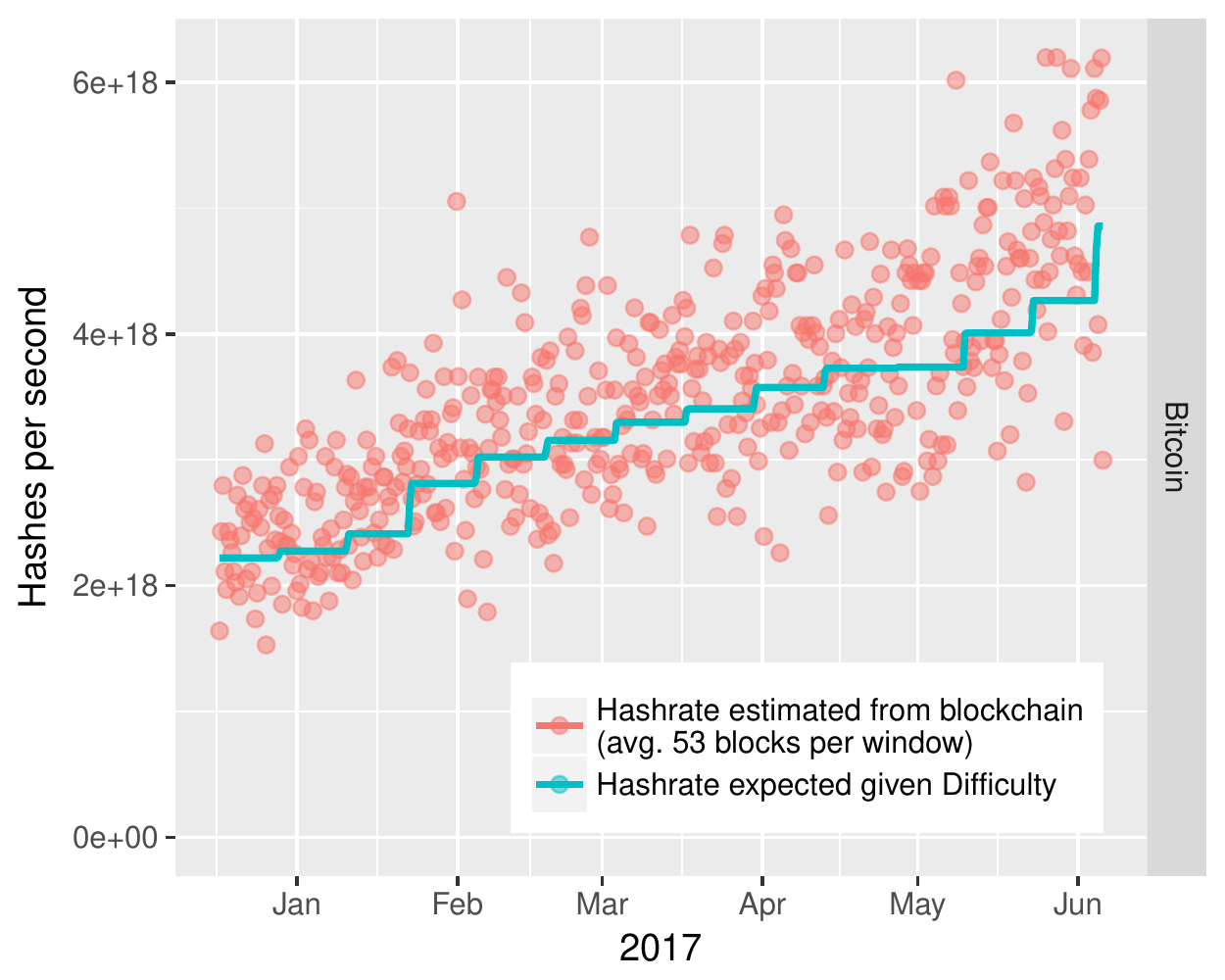} 
   \caption{Network-wide hash rate, computed using MoM, of the Bitcoin network compared to the hash rate calculated based on network difficulty and target value. Points represent a window of 500 minutes (about 50 blocks); for clarity, only a sample of all possible windows is plotted.}
   \label{fig:hashes-bitcoin}
\end{figure}

\begin{figure}[t] 
   \centering
   \includegraphics[width=.8\columnwidth]{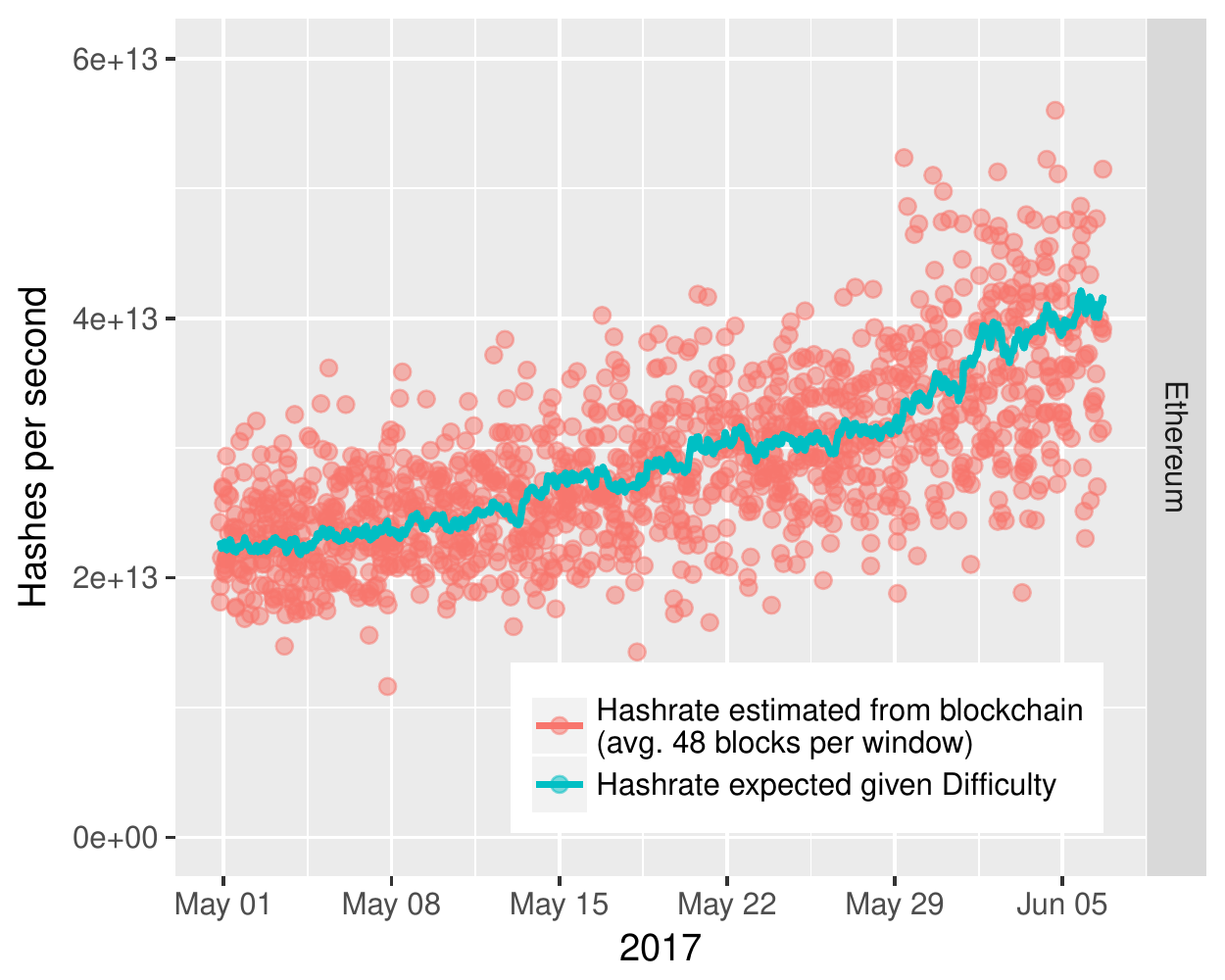} 
   \caption{Network-wide hash rate, computed using MoM, of the Ethereum network compared to the hash rate calculated based on network difficulty and target value. Points represent a window of 12.5 minutes (about 50 blocks); for clarity, only a sample of all possible windows is plotted.}
   \label{fig:hashes-ethereum}
\end{figure}

We also used our technique to estimate the number of hashes per block
at regular intervals in the Bitcoin and Ethereum networks.  As a comparison with a naive approach, we calculated the expected number of hashes based on the
difficulty, $D$, of blocks and the current target, $t$. For Bitcoin, this value is 
$E[\hat{h}]= \frac{2^{256}}{t\cdot600} =\frac{2^{32}D}{600}$; it is $E[\hat{h}]={2^{32}D}/{15}$ for Ethereum. 
The accuracy of our approach is apparent in
Figures~\ref{fig:hashes-bitcoin} and~\ref{fig:hashes-ethereum}.

\subsection{Hash Rate Estimation of a Subset of Miners}\label{sec:subset-miners}
We can estimate the hash rate of a single miner or a subset of miners with a simple extension of our technique from Section~\ref{sec:networkhash}. Specifically, for a given miner $m$, we
vary Eq. 5 such that the random variable
$\textbf{Y} = Y_1, \ldots, Y_n$ is defined as a function of
$\textbf{V}$ as follows:
\begin{equation}
Y_i = \textbf{1}_{V_i \leq t}(V_i) V_i = \left \{
\begin{array}{lr}
V_i & V_i \leq t \text{ and } \textbf{1}_m (V_i) \\
0 & \text{otherwise}
\end{array}
\right .
\label{eq:subset}
\end{equation}
where $\textbf{1}_m (x)$ is the indicator function equal to 1 when $x$
is mined by $m$ and 0 otherwise. Therefore, we construct
$\textbf{Y}$ only using the blocks issued by a given miner. The same approach also works for a subset of all miners. 

Additionally, the sample
mean, $\overline{Y}$, must also be adjusted such that it is the sum of
the observed hash values \textit{issued by the given miner}
divided by the number of intervals. Thus, once $\textbf{Y}$ and
$\overline{Y}$ only account for the blocks mined by a given miner,
Eq.~\ref{eqn:mm} and Alg. 1 are applicable.

\begin{figure}[t] 
   \centering
   \includegraphics[width=.8\columnwidth]{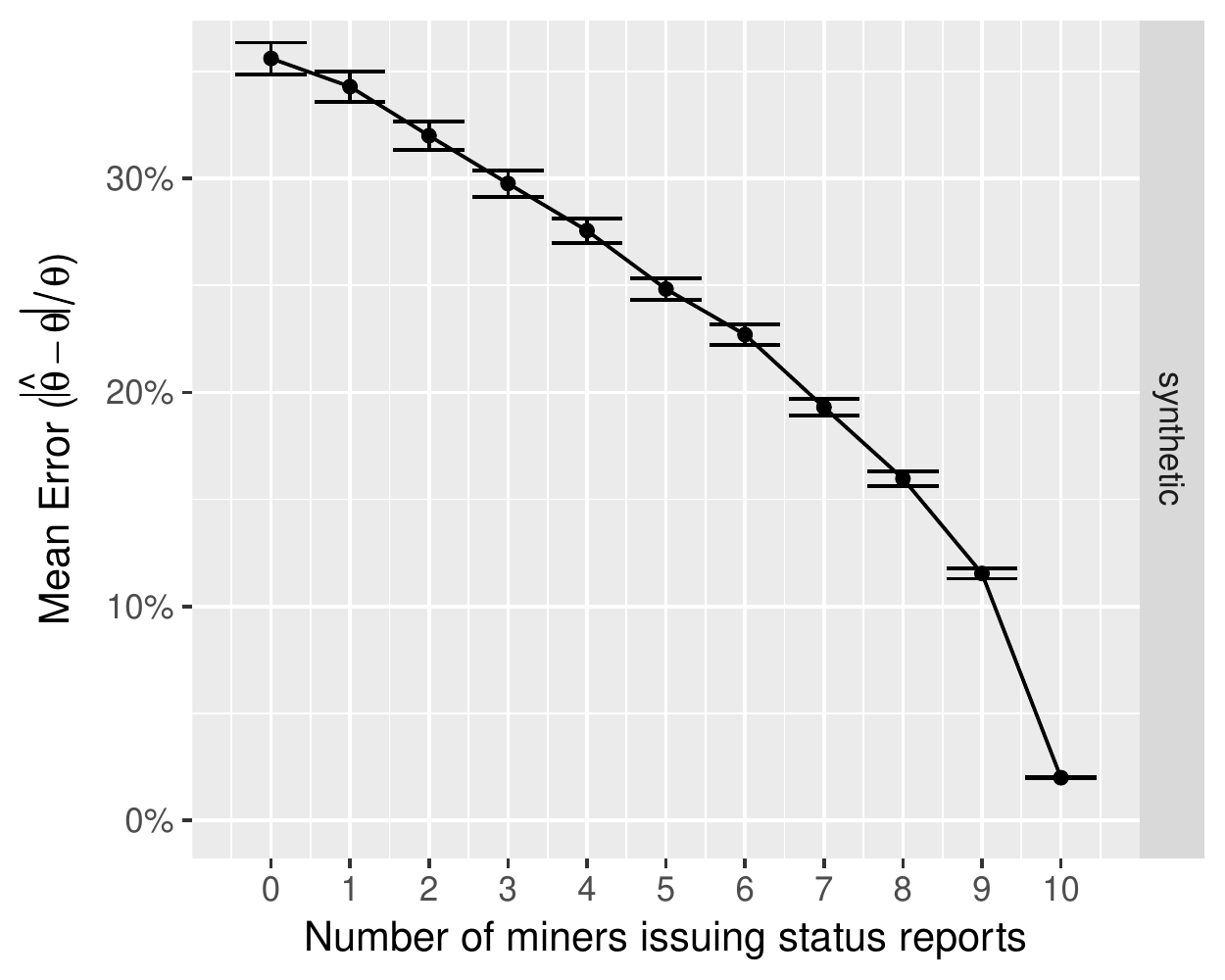} 
   \caption{Mean error of network hash rate estimation as the number of miners 
     issuing status reports increases. Synthetic blockchain with block discovery set to every 600 seconds. An 8-block window for MoM; status reports are issued about 320 times during the 8-block window. The accuracy of the network estimate increases dramatically as status reports are incrementally deployed. (Error bars are 95\% c.i.\ of the mean over 8,800 trials for each point.)}

   \label{fig:deployment}
\end{figure}

\para{Incremental Deployment.} If no miners adopt status
reports, network-wide hash rate can be calculated using the technique
detailed in Section~\ref{sec:networkhash}. For those miners who
deploy status reports, the estimator in
Section~\ref{sec:statusconsensus} can be used, while for those
remaining, the MoM estimator via Eq.~\ref{eq:subset} provides a solution; the two values are then summed.

We evaluated this approach using a synthetic block chain and 10 simulated miners, each with a fixed hash rate. Blocks were targeted for discovery every 600 simulated seconds. For hundreds of trials, we estimated the  hash rate of the network using only blocks. In subsequent trials, we allowed an increasing number of miners to issue status reports every 15 seconds. For the remaining miners that did not issue reports, we used  Eq.~\ref{eq:subset} to compute the number of hashes they
performed, i.e.,  we used only the blocks
issued by those miners not producing status reports. We then summed our
estimates to calculate network-wide hash rate.

Figure~\ref{fig:deployment} shows results of this experiment for a window of 5000 seconds (about 8 blocks, and about 320 status reports). Since 5000 seconds is a relatively small window for the MoM estimator (see Figure~\ref{fig:hashes-synthetic}), the accuracy is initially low at 35\%. The error in estimating the
hash rate decreases rapidly as  miners  adopt the  status
report mechanism.

\section{Tail Bounds}\label{sec:bounds}

As the CDFs plotted in Figures~\ref{fig:n-hat} and~\ref{fig:hashes-synthetic} show, our estimators have a distribution that is wide when there are too  few  reports or blocks, respectively, to work with.  In this section, we characterize these tails. We provide Chernoff bounds for our status-reports-based method and employ empirical bootstrapping for our blockchain-only method.

These tail bounds provide a defense for thwarting the efforts of attackers who seek to push falsified results to the estimators.  First, we describe the tail bounds, and then discuss their use against attackers. 

In Section~\ref{sec:implementation-analysis}, we use these bounds to both characterize the variance of our estimator in practice, and as a method to thwart certain attacks.

\subsection{Chernoff Bound for Status Reports}\label{sec:chernoff}
In Appendix~\ref{sec:sr_chenoff}, we derive the upper and lower tail
Chernoff bound on our estimate of $\beta$. The upper bound is the
following
\begin{eqnarray}
 P\left(\frac{\beta-\hat{\beta}}{\hat{\beta}} \geq \pi \right) \leq \texttt{exp}\left[ \frac{n \pi}{1 + \pi} - n \ln(1 + \pi) \right], \end{eqnarray}
where $n$ is the number of status reports in a window and $\pi$
represents the relative deviation of $\hat{\beta}$ from $\beta$. This
result gives us a natural interpretation of the Chernoff bound, and a
method to set it. In order to conservatively estimate a miner's mining
power, it must be the case that $\hat{\beta}$, estimated by
$\overline{V}$, in Equation~\ref{eqn:h_hat_m}, is large. A smaller
$\hat{\beta}$ implies that we are more likely to \textit{overestimate}
a miner's mining power. Therefore, we want the fractional change from
$\hat{\beta}$ to $\beta$ (described by
$(\beta-\hat{\beta}) / \hat{\beta}$) to be bounded. When this
fractional change is larger, our estimate for a miner's mining power
is \textit{higher}.  

A merchant would use the bound as follows. First, from $n$ status reports issued by a miner, she computes $\hat{\theta}_m$  using Eq.~\ref{eq:theta_hat_m}, and then sets $\hat{\beta}_m=S/\hat{\theta}_m.$ Given $n$, she finds the value of $\pi$ such that the RHS of Eq.\ref{eqn:h_hat_m} is less than or equal to a threshold (e.g., 0.05). She then assumes $\beta_{mL}=\hat{\beta}_m/(\pi+1)$ as a lower bound with high probability. We now have  an upper bound on 
the number of hashes performed by the miner as $\theta_{mH}=S/\beta_{mL}$.

In Section~\ref{sec:implementation-analysis}, we also make use of the lower bound with a similar formula, also derived in the Appendix: 
\begin{eqnarray}
P\left(\frac{\hat{\beta}-\beta}{\beta} \geq \pi \right)
 \leq \frac{1}{1+\pi} \texttt{exp}\left[-n(\pi - \ln(1 + \pi))\right].
\end{eqnarray}
In this case, given $n$ reports, the merchant solves for an appropriate $\pi$ that meets her threshold,  and then sets $\beta_{mH}=\hat{\beta}_m(\pi+1)$ and 
$\theta_{mL}=S/\beta_{mH}$.

\subsection{Empirical Bounds for MoM}\label{sec:MoM-bound}

We found that tail bounds for Eq.~\ref{eqn:mm} are not easily derived
using standard techniques, such as Chernoff or Chebychev. We therefore
calculate an empirical bound based on the well-known {\em
  bootstrap}~\cite{Efron:1982,Casella:2002} technique.
  
Specifically, given a window of intervals containing a sequence of $n$ blocks $\textbf{O}= O_1,\ldots,O_n$, we create 10,000 new samples, each created by selecting $n$ blocks {\em with replacement} from $\textbf{O}$. For each new sample, we compute its sample mean $\overline{Y}$. We then select the 5th percentile of this distribution as $\overline{Y}_L$ and solve for $\tilde\beta_{L}$ using Eq.~\ref{eqn:mm}. Finally, we have 
$\theta_{H}=S/\tilde\beta_{L}$.  Similarly, from the 95th percentile, we compute $\overline{Y}_H$ and $\tilde\beta_{H}$, as well as $\theta_{L}=S/\tilde\beta_{H}$. The approach is limited in accuracy for  windows containing only a handful of blocks.

\section{Implementation and Analysis}\label{sec:implementation-analysis}

 In this section, we apply our mining estimates to the problem of
quantifying the threat posed by double-spend attacks. Specifically, we
answer the question, {\em at what block depth can a merchant safely
  release goods to a consumer given current estimates of hash rates on
  the main chain?} We make use of historical data from the Bitcoin and
Ethereum blockchains to characterize this value in practice. We also
generated synthetic blockchain data so that we can evaluate the
accuracy of our approach against known miner power.

\subsection{Quantifying the Probability of a Double-spend Attack}

In Section~\ref{sec:ds}, we describe the double-spend attack against a transaction that appears in block $B_1$, which is a child of block $B_0$.
Using the techniques from previous sections, we can quantify the
amount of mining power devoted to mining on block $B_1$ and its
descendants. Specifically, an estimate of all miners working on $B_1$
and descendants is available from Eq.~\ref{eqn:h_hat} using the
blockchain-only MoM estimator; or the same estimate can be more accurately
calculated as the sum of individual miners' powers from status
reports, using Eq.~\ref{eqn:h_hat_m}.

Let $\theta_0$ be the network-wide hash rate in the case of MoM, or the
sum of hash rates for all miners in the case of status
reports, estimated from a {\em pre-window} of $w$ blocks {\em ending with}
$B_0$. Let $\theta_i$ be the   hash rate calculated from a {\em post-window}
of blocks {\em starting with} $B_0$ until $B_i$, where $i$ is the latest block on the chain descendent from $B_1$. The merchant assumes
that the amount of attacker mining power working against $B_1$ and its
descendants is the complement of the  proportion of mining power in the post-window to the pre-window:
\begin{equation}
q_i=1-\frac{\theta_i}{\theta_0}.\label{eq:q-i}
\end{equation}
Of course, past performance is no prediction of
what miners will do in the future: perhaps even after 1,000 blocks,
the attacker will attempt to double-spend. To account for this
possibility, we assume that from the honest miner's block $i+1$
onward, the attacker's mining power will be fixed at 12.7\% (see Section~\ref{sec:ds}). That is, we
use a revision of Nakamoto's formula:
\begin{align}
  \mathcal{D}(q_i,z)=\begin{cases} 
 1- \mathlarger{\sum}_{k=0}^{z} \frac{\lambda^k e^{-\lambda}}{k!}\left(1-(\frac{q*}{1-q*})^{z-k}\right)
   & \!\!\!\!\!\text{, if }  q_i<\frac12 \\ 
    1 & \!\!\!\!\! \text{, if } q_i\geq\frac12 \\
  \end{cases}\label{eq:satoshi-prime}
\end{align}
where $\lambda=\frac{zq_i}{1-q_i}$ and $q*=0.127$.
If the resulting probability is below the merchant's threshold (e.g., 0.1\%), the goods are released to the consumer. 
For each block added to the chain, the merchant re-evaluates.

\para{Bounding risk.} A more conservative merchant, with perhaps more coin at risk, will account for  error in the estimates. For status
reports, the merchant can  make use of Chernoff bounds from Section~\ref{sec:chernoff}; and for  the MoM
estimator, the merchant can use bootstrapped estimates of the sample's tail
percentile from Section~\ref{sec:MoM-bound}. The same approach can be used to account for attackers, as we detail below in Section~\ref{sec:attacks-on}.

\begin{figure}[t] 
\centerline{\includegraphics[width=.7\columnwidth]{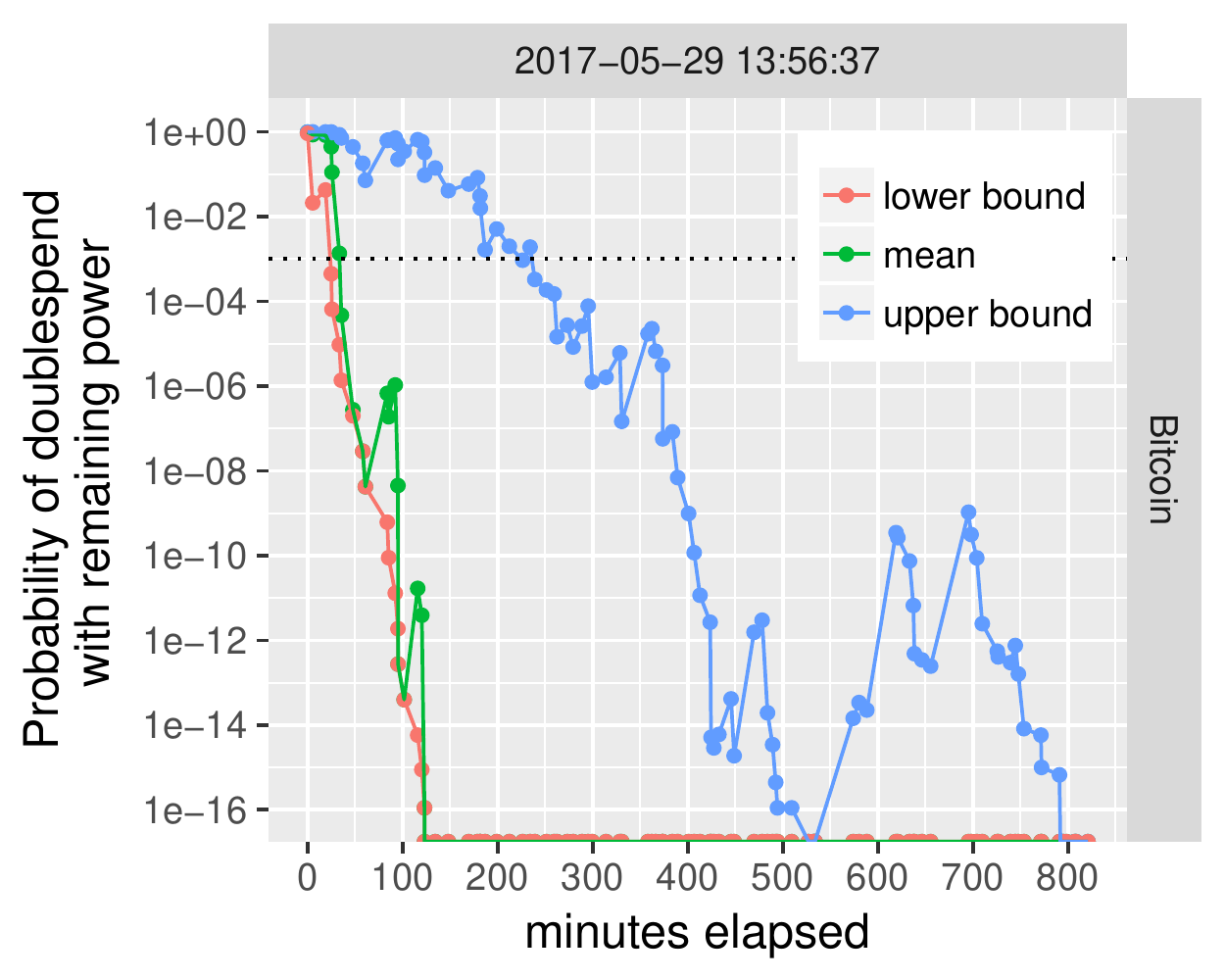} }
\centerline{\includegraphics[width=.7\columnwidth]{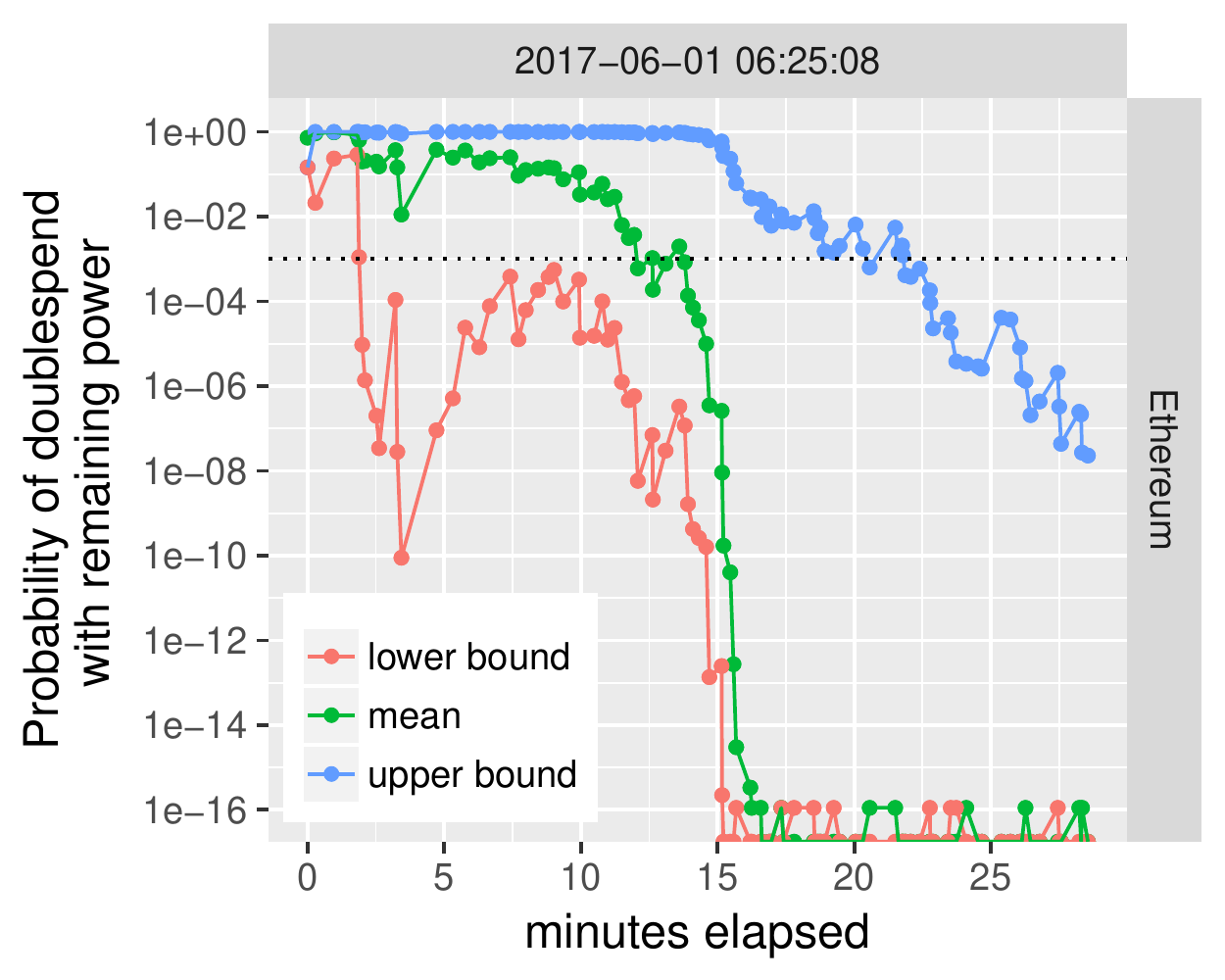} }
\caption{Results of applying Eq.~\ref{eq:satoshi-prime} to an example block
  as its depth increases. The upper and lower bounds are calculated
  using bootstrapped estimates of the sample's 5th  and 95th   percentile, respectively (Eq.~\ref{eq:q-iL} and~\ref{eq:q-iH}).}
\label{fig:depth-ex-prob} 
\end{figure}

To consider the worst-case
scenario, we use an upper-bound estimate during the pre-window, and a lower-bound estimate during the post-window.
\begin{equation}
q_{iL}=1-\frac{\theta_{iL}}{\theta_{0H}}.\label{eq:q-iL}
\end{equation}
The consequence is that goods are released later. 

A best-case scenario for an impatient merchant seeking the earliest time to release his goods is:
\begin{equation}
q_{iH}=1-\frac{\theta_{iH}}{\theta_{0L}}.\label{eq:q-iH}
\end{equation}
In other words, the two bounds characterize the error of estimating attacker mining power using $q_i$ (Eq.~\ref{eq:q-i}). Both bounds can be applied to Eq.~\ref{eq:satoshi-prime}.

\para{Implementation.} 
We implemented our techniques for Bitcoin and Ethereum,
using only information available on the blockchain.  We released a 
public demo for Bitcoin at \url{http://cs.umass.edu/~brian/blockchain.html}.
Once mining pools
begin to release status reports, we will  update to increase the accuracy of our estimates, as discussed in Section~\ref{sec:subset-miners}.  The site is meant to show that we do not require
updates to the existent protocols or underlying topology. (We will  release source code at camera ready.)

\para{Example output.}
Figure~\ref{fig:depth-ex-prob} is  example of our technique applied to a single block as it ages. The figure shows the probability of attacker success (green line computed using Eq.~\ref{eq:satoshi-prime}) given block depth in terms of minutes, computed using a pre-window of 100 blocks. We also calculated the upper and lower bounds on success; the red and blue lines show Eqs.~\ref{eq:q-iL} and~\ref{eq:q-iH}, respectively. 
To highlight a difference between Bitcoin and 
Ethereum, the figure is in minutes, instead of the number of blocks in  the post window. Below, we examine the historical performance of the two networks.

\begin{figure}[t] 
   \centering
   \includegraphics[width=.85\columnwidth]{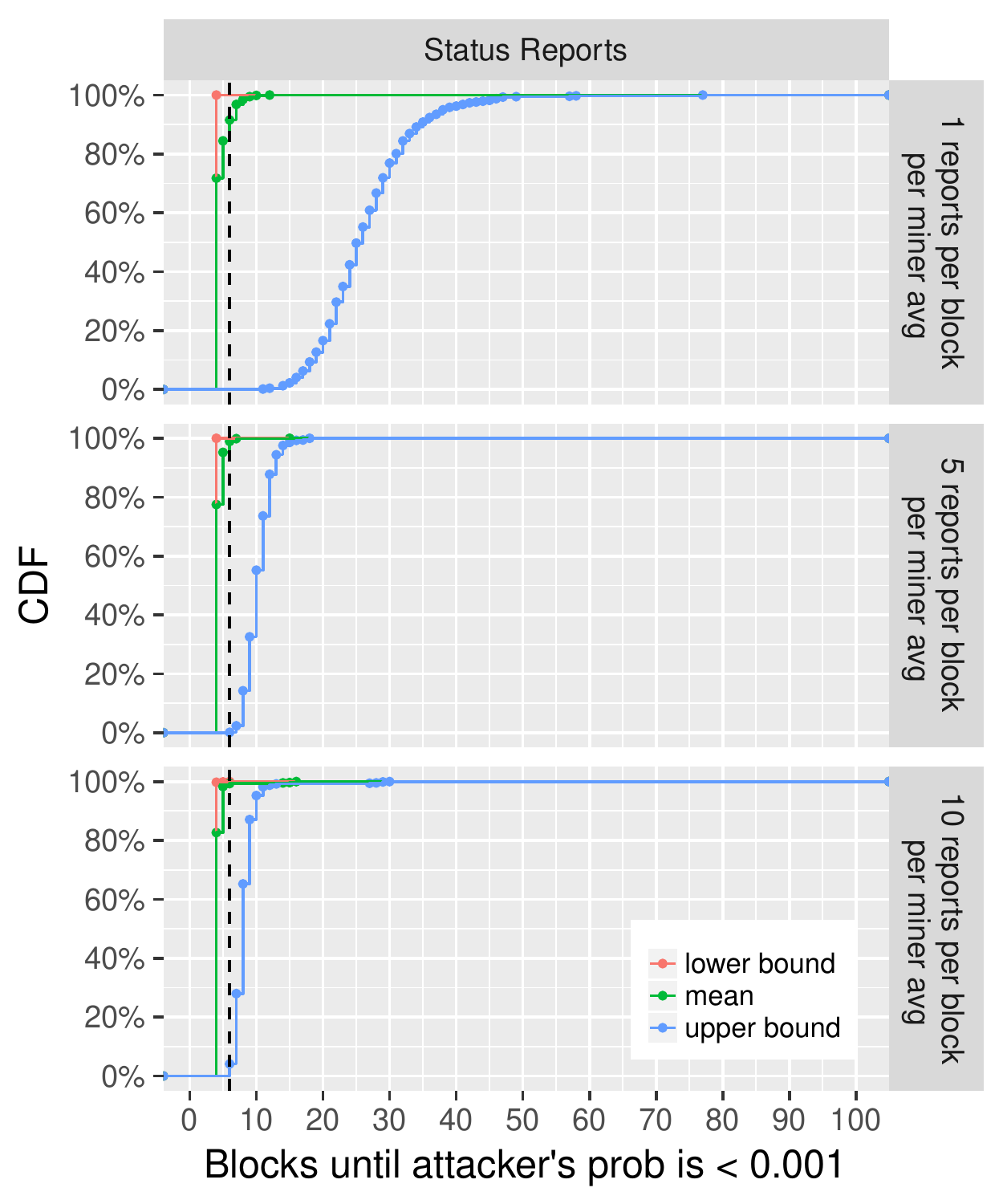}
   \caption{CDF for the depth required to defeat attackers with a
     given probability for synthetic data. The Chernoff bounds are
     weaker than the empirical bootstrapping method used for real
     blockchain data, but still demonstrate better results in the worst case.}
   \label{fig:depth-syn}
\end{figure}

\subsection{Performance of Status Reports on Synthetic Data}
We implemented our techniques for status reports on a synthetic blockchain to quantify its performance. The blockchain was parameterized to issue blocks about once every 600 simulated seconds. All miners issued status reports at rate of either 1, 5, or 10 times per block. 

 We then sampled several hundred blocks from the chain and computed Eqs.~\ref{eq:satoshi-prime},~\ref{eq:q-iL}, and  \ref{eq:q-iH} using estimates from status reports and Chernoff bounds for each sample block as its depth increased. We used pre-windows of 100 blocks. We logged  when each  equation reached a set a threshold of 0.1\% probability of attacker success.
Figure~\ref{fig:depth-syn} shows these results as a CDF for each equation. 

The results show that when each mining pool issues on average 1 report between block announcements, at most 6 blocks are required 90\% of the time to reach the 0.1\% attack success threshold. This translates to one report every 15 seconds by each Ethereum mining pool, or one report every ten minutes for each Bitcoin mining pool. However, at this rate, the upper bound can be high. By increasing reports to about 10 per block per miner (about 2 every 3 seconds in Ethereum, and one every minute in Bitcoin), even the upper bound shows that 99\% of blocks are within a depth of 13.

\subsection{Historical Bitcoin and Ethereum Performance}

\begin{figure}[t] 
   \centering
   \includegraphics[width=.75\columnwidth]{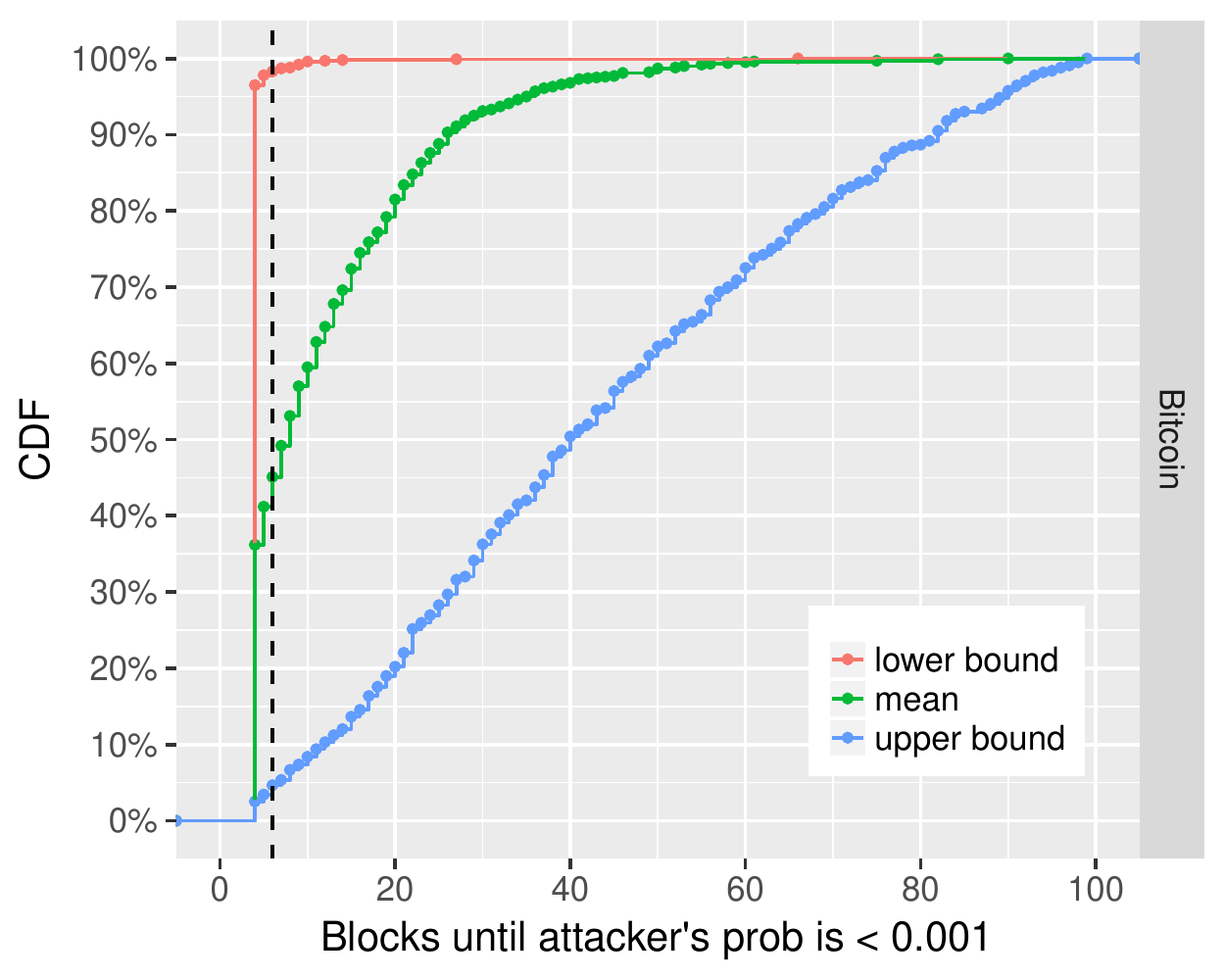}
   \includegraphics[width=.75\columnwidth]{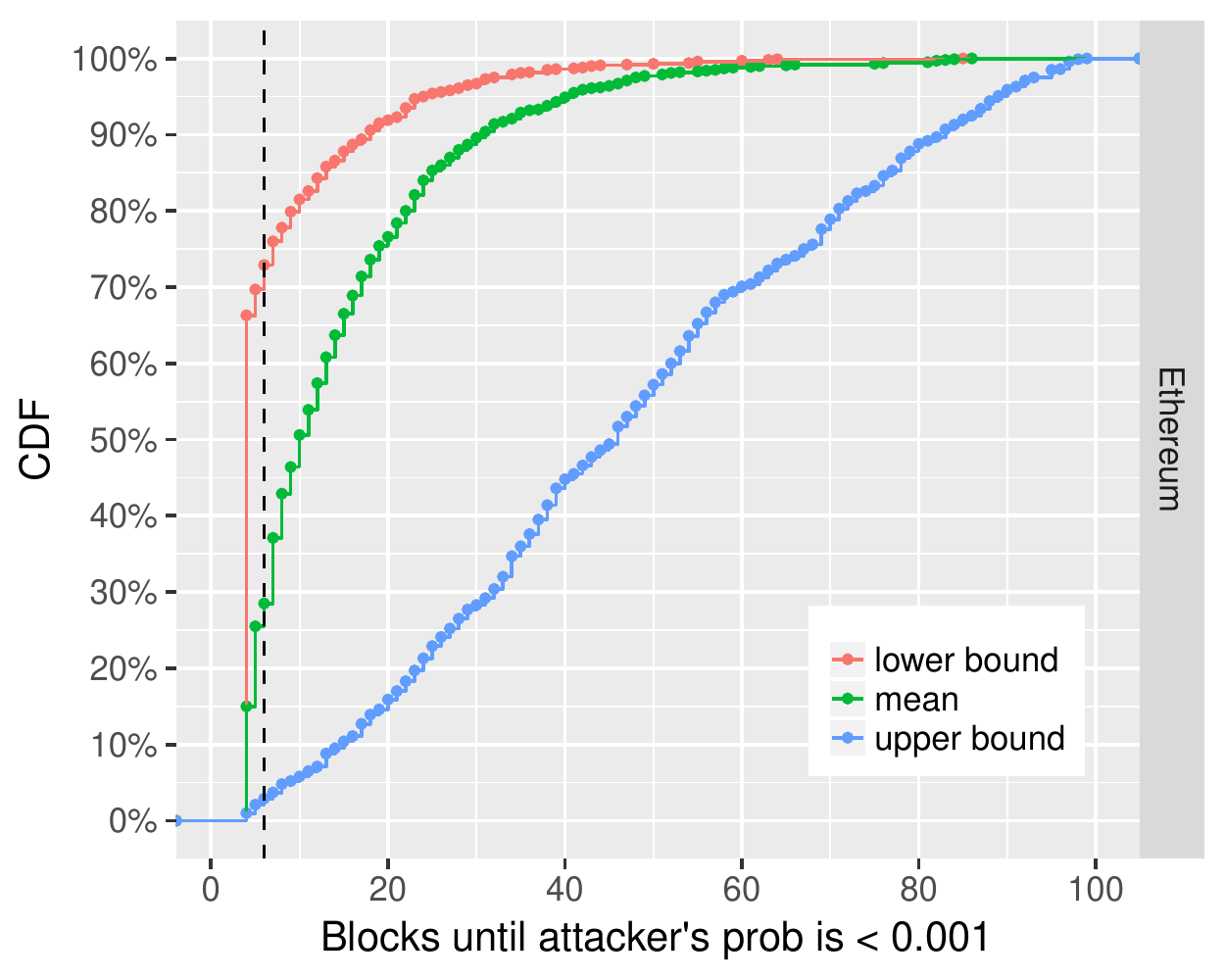}
   \caption{CDF for the depth required to defeat attackers with a
     given probability. Because Ethereum includes ommers to the
     blockchain, we are able to provide a tighter estimate. }
   \label{fig:depth}
\end{figure}

We then performed the same analysis for historical data from Bitcoin and Ethereum. We computed values for Eqs.~\ref{eq:satoshi-prime},~\ref{eq:q-iL}, and  \ref{eq:q-iH} for over 2000 randomly selected blocks, each with a 100-block pre-window. For each block, we determined the depth required  such that the probability of attacker success was less than 0.1\%.  

Figure~\ref{fig:depth} shows the CDF of these experiments for the two networks. First, the results are much poorer than the status report method. Second, it is notable that the two systems exhibit similar mean performance.  For both  systems, to ensure the probability of attacker success is lower than 0.1\%, at least 10 blocks are required about 40\% of the time; the upper bounds are much weaker, requiring at least 40 blocks 50\% of the time. These values vary, but a strength of our technique is that it is
calculated by merchants for specific blocks.
   
Ethereum's 15-second inter-block time results in a greater number of abandoned blocks and forks (called {\em ommers}); these are recorded to the blockchain as part the GHOST algorithm. Our algorithm makes use of the ommers as part of its estimate, and thus the bounds are tighter since there is more information in each window. While Figure~\ref{fig:depth-ex-prob} demonstrates that Ethereum more quickly reduces the risk of a double-spend attack in terms of wall clock time, Figure~\ref{fig:depth} shows that in terms of block depth, the two networks offer equivalent security because they implement the same algorithm.

The dashed vertical line represents bitcoind's choice of 6 blocks for declaring blocks as confirmed. According to the mean  estimate, this choice is too early about 70\% of the time in both networks. While there  is error in the MoM technique, the experiment demonstrates that the 6-block choice is conservatively too early at least 30\% of the time for Ethereum, according to our upper-bound estimate from Eq.~\ref{eq:q-iL}. Because there are few forks in the Bitcoin blockchain, there is less data for making estimates, and we cannot make the same conservative claim for Bitcoin but expect it holds true.

\subsection{Bandwidth Costs}\label{sec:bandwidth}
Using only information available from the blockchain to estimate mining power requires no additional bandwidth.

The bandwidth required by status reports depends on the number and
rate at which they are issued. The size of a status report is equal to
the block header, plus a mining pool identifier and cryptographic
signature.  Status reports would be 80 bytes, the same size as
block headers for Bitcoin; similarly, they would be 508
bytes for Ethereum. We assume that identifiers and signatures can be amortized by
an SSL connection to a web site.  Each report also can be easily encoded as
1--2 tweets on Twitter, which provides validated accounts and a secure
global infrastructure for distributing short announcements. (It is unclear if this approach would violate Twitter's terms of service.) Secured
RSS/json feeds from websites could also be used. In the case that two pools claim the same block, the claims can easily be resolved by the real owner's
signing of the claim with the block's coinbase private key. Note that bandwidth costs are independent of block size   since they depend on only the block header.

To make a comparison  of costs fair, we consider the frequency of status reports in terms of the network's block interval: e.g., 10 per 600 seconds in Bitcoin, and 10 per 15 seconds in Ethereum. We assume that 20 mining pools are active on each network, which mirrors current conditions. For Bitcoin, if ten 80-byte status reports are sent per miner per 600 seconds, then the total cost for recipients is 
$0.026~\mbox{KBps}$. This additional traffic is small compared to keeping track of the blockchain itself, which is about 1~MB/600~sec = 1.7KBps.

For Ethereum, if ten 508-byte status reports are sent per miner per 15 seconds, then the total cost for recipients is 6.6~KBps.  This additional traffic is much higher  than  keeping track of the blockchain itself, which is about 10KB / 15sec = 0.6~KBps. In absolute terms, this is still low --- for example, streaming a song from Spotify at the lowest quality setting is about 12~KBps. Further, Ethereum performance would benefit significantly from just 1 report per miner per 15 seconds (i.e., costing 0.66~KBps).

\subsection{Attacks on Hash Rate Estimates}
\label{sec:attacks-on}
Malicious miners may try to issue status reports that cause our hash rate estimates to be falsely higher or lower. 
In particular, a miner may want his mining rate to appear higher during the post-window, or lower during the pre-window.

\para{Attacker Model.} We assume our attacker has a total hash rate that is applied to the main chain in full during the pre-window, and then partially during the post-window. Outside of this restriction, attackers are challenging to model.  For example, an attacker may take advantage in a  change between the pre- and post-window in exchange rate of coin to a fiat currency that is used to pay electricity. Or an attacker may similarly take advantage of spot instances that fall in price on a cloud service, such as EC2. We do not take into account such economics or externalities. We discuss this limitation further in Section~\ref{sec:limitations}.

It is easy for an attacker to falsely lower their pre-window mining rate by simply not mining. Such an attacker is outside our model --- but to put it another way, we assume the pre-window duration is sufficiently long to dissuade attackers from giving up income from not mining. 

\para{Attack analyses.}  
An attacker might attempt to pre-mine the nonce values. However, the reports, just like blocks, are tied to a particular prior block, preventing the miner from creating status reports before a block is mined. 

Similarly, an attacker might attempt to  pre-mine within the window of a single block. For example, they may report the second  order statistic as the second report. To prevent this attack, we can require status reports to include a nonce based on the previous status report. For example, if the report header contains a nonce $n_i$, then the nonce used for POW is the hashed concatenation: $n'_i=H(n'_{i-1}|n_i)$, where $n_0$ is the hash of the prior block and $H(\cdot)$ is a hash function.

A more advanced strategy is for an attacker to wait until she gets lucky. She mines until her status reports produce an overestimate, and then diverts resources to a double-spend attack. Or, the attacker 
 stops mining towards a status report at each interval when finding a minimum hash that satisfies the
 target mean; for the remainder of the interval, an
 attacker uses her mining power to execute a double-spend attack. 

For these more advanced strategies, the attacker is taking advantage of the tail of the estimator's  distribution. Chernoff bounds are a powerful technique against such attacks. 
Yet, we do not claim that Chernoff bounds defend against all possible attacks. Consider an attacker with mining power $x$ that devotes $2x/3$ to mining honestly, and $x/3$ to the double-spend attack. If the miner never reveals their $x/3$ power until the attack succeeds, our techniques would certainly fail to detect this attacker. We have not analyzed whether this attack is economically profitable. 

In the case of status reports, an attacker can lower the merchant's estimate of their mining power by skipping status reports or sending a hash that is not their lowest. We hypothesize that this attack would be detectable because the rate estimated by the status reports and that from the blocks they add to the chain (see Section~\ref{sec:subset-miners}) would not be statistically equal. We do not expect this comparison to work for short windows of time, but neither have we determined the time necessary for any differences to be statistically significant.

In the case of blockchain-only MoM estimations, the bootstrap sample distribution provides protection similar to the Chernoff bounds as the tails are difficult for an attacker to avoid. Notably, attackers of the blockchain-only method face an additional hurdle: to affect the outcome, they must mine valid blocks.

\section{Limitations}
\label{sec:limitations}
There are several limitations to our approach and contributions. 
As discussed in Section~\ref{sec:attacks-on}, Bitcoin and Ethereum are open blockchains that allow miners to join or leave at any time~\cite{Vukolic:2015}. We are unable to account for latecomers that were previously silent during the pre-window or new to mining starting at the post-window. 

Real attackers may employ strategies that are more complex than we considered. For example, mining pools may hide their hash rates by switching their mining resources between networks with the same  proof of work algorithm, such as between Bitcoin and Litecoin (Ethereum's POW algorithm is designed to advantage GPUs rather than ASICs). We note therefore, that a conservative merchant may elect to sum the hash rates across a series of blockchains. (Fortunately, our approach is not subject to the Sybil attack~\cite{Douceur:2002}, as we do not separate out mining pool workloads.) But in general, externalities that we cannot observe or quantify may drive attacker strategy and  behavior.  None of our analysis considers the economic profitability of attacker strategy. 

We have not studied the optimal duration of the pre-window in Section~\ref{sec:implementation-analysis}. A longer window assumes steady hash rates over long periods of time, while missing finer-grained fluctuations. A short window accounts for recent history while missing long-term trends. 
Additionally, in the same section, we do not offer a specific threshold for 
bootstrap or Chernoff bounds. On the other hand, this may be considered a parameter set by the merchant, who may tune the tradeoff between security and delay to their preference.

Finally, we offer no direct incentive for miners to issue status reports. However, doing so may encourage greater trust and use of  blockchains, which benefits  miners.
\section{Related Work}

There have been many informal proposals to 
determine the amount of 
hash power that went into unconfirmed transactions --- see \cite{Bishop:2015} for a list. To the best of our  knowledge, no algorithms have been formally defined, evaluated, or implemented; these suggestions are so preliminary that we cannot compare our approach. Further, none suggest a method that uses only existing blockchain information. (A preliminary version of this paper appeared as a tech report~\cite{Ozisik:2016}.)

The security of a 6-block wait for transaction confirmation has been 
studied by Rosenfeld~\cite{Rosenfeld:2012} and discussed by Bonneau~\cite{Bonneau:2015a}; see also~\cite{bitcoin:confirmation}. Many papers have examined the double-spend attack in a variety of contexts. Sompolinsky introduced the GHOST protocol~\cite{Sompolinsky:2015}, now incorporated in Ethereum. They showed that double-spend attacks become more effective as either the block size or block creation rate increase (when GHOST is not used).   Sapirshtein et al.~\cite{Sapirshtein:2015} first observed that some double-spend attacks can be carried out essentially cost-free in the presence of a concurrent selfish mining~\cite{eyal:2014} attack. More recent work extends the scope of double-spends that can benefit from selfish mining to cases where the attacker is capable of \emph{pre-mining} blocks on a secret branch at little or no opportunity cost~\cite{Sompolinsky:2016} and possibly also under a concurrent eclipse attack~\cite{Gervais:2016}. 

Several past work have relied on stability in the blockchain as a requirement for a higher-level service. For example, sidechains\cite{Back:2014} employ a confirmation period, which is ``a duration for which a coin must be locked on the parent chain before it can be transferred to the sidechain,'' stating that ``a typical confirmation period would be on the order of a day or two'' but not reasoning why. Our technique offers a quantitative approach to selecting the confirmation period. Lightning networks~\cite{bitcoin:lightning, Heilman:2017} and fair-exchange protocols~\cite{Barber:2012} similarly assume that an initial commitment or refund  transaction is not revokable (via a double-spend attack), and our approach would quantify that risk.

\section{Conclusion}\label{sec:conc}
We  designed and evaluated two methods to accurately estimate
network hash rates to quantify the probabilistic consensus of blockchain
systems. Our first method is based on short status reports issued by
miners, while the second uses only  blocks that are published to the
blockchain. The latter approach is less accurate than status reports
because there is less information available per window of time. To evaluate the accuracy of
both methods, we derived bounds on our estimates and presented
simulations using a synthetic blockchain as ground truth. We also
showed that these methods can be used together in an incremental deployment strategy.

We also implemented our blockchain-only estimator to show the
historical network-wide hash rate of Ethereum and Bitcoin. Finally, we provided a framework to
estimate consensus on blocks and transactions, using our estimates. On
Bitcoin and Ethereum, we analyzed the depth required before an
attacker's success is estimated to be 0.001 or less. We emphasized the
importance of status reports by demonstrating, using synthetic blockchain data, that this depth is
significantly reduced.


\appendix

\section{Status Report Chernoff Bounds}\label{sec:sr_chenoff}
We derive a bound on the relative deviation of
$\hat{\beta}$ from $\beta$.

\para{Upper tail bound.}
Let $\textbf{R} = \sum_{i=1}^n V_i$
with $V_i \sim \texttt{Expon}(\beta)$. Jansen~\cite{Janson:2014} shows  that for any $\lambda \leq 1$,   \vspace{-.5ex}
\begin{equation}
P(\textbf{R} \leq \lambda E[\textbf{R}]) \leq \texttt{exp}\left[{-\frac{1}{\beta} E[\textbf{R}] (\lambda - 1 -\ln(\lambda))}\right].
\end{equation}

\noindent In our case, $E[\textbf{R}] = n \beta$ and $\textbf{R} = n \hat{\beta}$, where $\hat{\beta}$ is the observed sample mean of the status reports. Let $\lambda = 1 / (1 + \pi)$. It follows that
\begin{eqnarray}
\label{eq:beta_large_CB}
P\left({\beta-\hat{\beta}}/{\hat{\beta}} \geq \pi \right) & = & P(n \hat{\beta} \leq \lambda n \beta) \nonumber\\
& \leq & \texttt{exp}\left[{-\frac{1}{\beta} n \beta (\lambda - 1 -\ln(\lambda))}\right] \nonumber\\
& = & \texttt{exp}\left[ \frac{n \pi}{1 + \pi} - n \ln(1 + \pi) \right].
\end{eqnarray}

\para{Lower tail bound.}
Jansen~\cite{Janson:2014}  shows that for any $\lambda \geq 1$,
\begin{equation}
\!\!\!\!\!P(\textbf{R} \geq \lambda E[\textbf{R}]) \leq \lambda^{-1} \texttt{exp}\left[{-\frac{1}{\beta} E[\textbf{R}] (\lambda - 1 -\ln(\lambda))}\right]
\end{equation}
If $\lambda = 1+\pi$, then we have
\begin{eqnarray}
\label{eq:beta_small_CB}
P\left({\hat{\beta}-\beta}/{\beta} \geq \pi \right)\!\!\!\! & = &\!\!\!\!\! P(n \hat{\beta} \geq \lambda n \beta) \nonumber\\
\!\!\!\!& \leq &\!\!\!\!\! \lambda^{-1} \texttt{exp}\left[{-\frac{1}{\beta} n \beta (\lambda - 1 -\ln(\lambda))}\right] \nonumber\\
\!\!\!\!& = &\!\!\!\!\! \frac{1}{1+\pi} \texttt{exp}\left[-n(\pi - \ln(1 + \pi))\right].
\end{eqnarray}

\bibliographystyle{ACM-Reference-Format}
\bibliography{references}

\end{document}